\documentclass[12pt]{article}

\usepackage{amsfonts}
\newcommand{\p}[1]{(\ref{#1})}

\newcommand{\be}{\begin{equation}}
\newcommand{\bea}{\begin{eqnarray}}
\newcommand{\ee}{\end{equation}} \newcommand{\eea}{\end{eqnarray}}

\begin{document}
\thispagestyle{empty}

\vspace{2cm}

\begin{center}
{\bf On the Time Dependent Oscillator and the Nonlinear
Realizations
of the Virasoro Group }\vspace{0.3cm} \\

V.P. Akulov${}^{a,b}$ \footnote{E-mail:
akulov@gursey.baruch.cuny.edu}, Sultan Catto${}^{b,c}$
\footnote{E-mail: scatto@gc.cuny.edu}, Oktay
Cebecio\v{g}lu${}^{d}$ \footnote{E-mail:
ocebecio\v{g}lu@yahoo.com} and A. Pashnev${}^{e}$
\footnote{E-mail:
pashnev@thsun1.jinr.ru}\\
\vspace{0.3cm}

${}^a${\it Department of Natural Sciences, Baruch
College of the City University of New York}\\
{\it New York, NY 10010, USA}\\
\vspace{0.3cm}

${}^b${\it  Physics Department\\
The Graduate School and
University Center\\
The City University
of New York,}\\
{\it 365 Fifth Ave, New York, NY 10016-4309,USA} \vspace{.3cm}\\

${}^c${\it Center for Theoretical Physics\\
The Rockefeller University}\\
{\it 1230 York Ave, New York, NY 10021-6399, USA} \vspace{.3cm}\\

${}^d${\it Department of Physics,
University of Kocaeli }\\
{\it Atat\"urk Bulvar{\i}, An{\i}tpark Yan{\i},
41300, Kocaeli, Turkey} \vspace{.3cm}\\

${}^e${\it Bogoliubov Laboratory of Theoretical Physics, JINR} \\
{\it Dubna, 141980, Russia}\vspace{.5cm}\\

{\bf Abstract}
\end{center}
\begin{center}
{\begin{minipage}{4.2truein}
                 \footnotesize
                 \parindent=0pt

Using the nonlinear realizations of the Virasoro group we
construct the action of the Conformal Quantum Mechanics (CQM) with
additional harmonic potential. We show that $SL(2,R)$ invariance
group of this action is nontrivially embedded in the
reparametrization group of the time which is isomorphic to the
centerless Virasoro group. We generalize the consideration to the
Ermakov systems and construct the action for the time dependent
oscillator. Its symmetry group is also the $SL(2,R) \sim SU(1,1)$
group embedded in the Virasoro group in a more complicated way.

\end{minipage}}\end{center}
                 \vskip 2em \par

\setcounter{footnote}0\setcounter{equation}0 \vfill
\setcounter{page}0 \renewcommand{\thefootnote}{\arabic{footnote}}

\newpage
\setcounter{equation}0
\section{Introduction}

The Time Dependent Oscillator ( so called Ermakov system[1]) has
been the subject of vigorous research for decades because of its
relevance to a large variety of physical phenomena, especially due
to its elegant mathematical properties and application potential
of its invariant. The list is very exhaustive which include in
particular the study of soliton solutions of nonlinear evolution
equations\cite{ioffe}, construction of time dependent integrals of
motion for the parametric harmonic oscillator used for the
canonical formulation of more general parametric systems and their
semiclassical quantization\cite{lewis}, theory of coherent and
squeezed states\cite{manko}, Berry's phase\cite{monte}, Noether's
theorem and Noether and Lie symmetries of the time dependent
Kepler system\cite{ray}, anisotropic Bose-Einstein condensates and
completely integrable dynamical systems\cite{haas}, cosmological
particle creation\cite{ray4}, scalar field cosmologies and
inflationary scenarios\cite{hawkins}, nonlinear
optics\cite{gonch}, propagation of water waves\cite{athor}, exact
solution for the Calogero system\cite{calogero1} and non-central
potential with dynamic symmetry\cite{haas2}, study of motions in a
Paul trap\cite{paul}, quantum mechanical description of highly
cooled atoms\cite{died}, emergence of non classical optical states
of light due to time-dependent dielectric constant\cite{agarwal},
particle distribution for beam in electric field\cite{drivotin},
nonlinear elasticity\cite{shahinpoor}, molecular
structures\cite{gaididei}, quantum field theory in curved
spaces\cite{finelly}, quantum cosmology\cite{rosu}, the connection
of the quantum time dependent oscillator to the free motion \cite{KC}
etc., among
others.

A vital modification of the Time Dependent Oscillator includes an
additional term in the potential proportional to the inverse
square of the coordinate - it is often referred to as the
anharmonic oscillator. This extra term is conformally invariant.
The analogous Conformal Quantum Mechanics (CQM) was investigated
in detail by De Alfaro, Fubini and Furlan\cite{AFF}. It was shown
in their paper that the consistent quantum treatment of the model
assumes the transition to the new time coordinate, which
transpires to be equivalent to the introduction of additional
oscillator-like term with a constant frequency in the potential.
Therefore the emerging physical Hamiltonian represents the
anharmonic oscillator with time independent frequency $\omega$.

Another fascinating feature of Conformal Quantum Mechanics
(CQM)\cite{AFF}, as well as its supersymmetric generalization --
SCQM \cite{AP}-\cite{G}, is the fact that they are the simplest
theories that allow the cultivation and development of methods for
investigation of more complicated higher dimensional field
theories. One should also note that in spite of its simplicity,
SCQM describes the physical objects such as a particle near the
horizon of black hole\cite{CDKKTP}, etc. The extended SCQM is also
closely related to the Calogero model with spin, which has
numerous physical applications.

The most adequate approach for understanding the geometrical
meaning of CQM and SCQM is the method of nonlinear realizations of
the symmetry groups underlying both theories - the group $SL(2,R)$
and its supersymmetrical generalizations $SU(1,1|1)$ and
$SU(1,1|2)$ respectively\cite{IKL1},\cite{IKL2}. In spite of its
power, this method does not allow the possibility of including in
the Hamiltonian of the theory the oscillator-like potentials
introduced in \cite{AFF}. As will be shown in this paper the
explanation for this lies entirely in the fact that in the
presence of the oscillator-like term the invariance group of the
action, though being the Conformal Group, is realized by the more
complicated transformations. These transformations for the
constant $\omega$, as well as for the time-dependent one (Ermakov
system), can naturally be embedded in the reparametrization group
of the time variable which is isomorphic to the centerless
Virasoro group. This embedding is rather nontrivial in the case of
nonvanishing $\omega$.

The structure of the paper is as follows. In Section 2 we  apply
the nonlinear realizations method to the Virasoro group and its
three dimensional subgroup $SL(2,R)$. We calculate the
transformation laws for parameters of these groups and construct
differential Cartan's Omega forms invariant under  these
transformations. They are then used for the construction of the
action for Conformal Quantum Mechanics in the Subsection 3.1. In
the Subsection 3.2 we illustrate the mechanism for the appearance
of the oscillator-like terms in the Omega-forms, and
correspondingly in the action. We show how the symmetry group of
this action, $SL(2,R)$, is non-trivially embedded in the Virasoro
group. In Subsection 3.3 we generalize these results to the
Ermakov systems with time dependent oscillator frequency. We
describe also the transformations of the time and phase space
variables which connect with each other the Hamiltonians with
different values of the harmonic oscillator frequency, including
the free motion and Ermakov systems. Some further anticipations of
the formalism we developed are included in conclusions.

\setcounter{equation}0
\section{The Nonlinear Realization of the Reparametrization Group}
The generators of the infinitedimensional reparametrization
(diffeomorphisms) group on the line parametrized by some parameter
$s$ are $L_m=is^{m+1}\frac{d}{d s}$ and form the Virasoro algebra
without central charge \begin{equation}\label{Virasoro}
\left[L_n,L_m\right]=-i(n-m)L_{n+m}.
\end{equation}
If one restricts to the regular at the origin $s=0$
transformations, it is convenient to parametrize the Virasoro
group element as\cite{IK1,IK2}
\begin{eqnarray}\label{coset1}
 &&G=e^{i\tau L_{-1}} \cdot e^{ix_1L_1} \cdot e^{ix_2L_2} \cdot
 e^{ix_3L_3}\ldots e^{i{x_0}L_0},
 \end{eqnarray}
 where all multipliers, except the last one, are arranged by the
 conformal weight of the generators in the exponents.

The transformation laws of the group parameters $\tau, x_n$ in
(\ref{coset1}) under the left action
\begin{equation}\label{left1}
G'=(1+ia)G,
\end{equation}
where infinitesimal element $a$ belongs to the Virasoro algebra
\begin{equation}\label{epsilon}
a = a_0 L_{-1}+{a_{1}}L_0+ {a_{2}}L_1+...+{a_{m}}L_m+... =
\sum_{n=0}^{\infty}a_nL_{n-1},
\end{equation}
are
\begin{eqnarray}\label{Transtau}
\delta \tau &=&a(\tau),\\
\delta x_0&=&\dot{a}(\tau),\label{Trans0}\\
\delta {x_1}&=& -\dot{a}(\tau)x_1+
\frac{1}{2}\ddot{a}(\tau),\label{Trans1}\\
\delta {x_2}&=&
-2\dot{a}(\tau)x_2+
\frac{1}{6}\stackrel{\ldots}{a}(\tau),\label{Trans2}\\
&& \ldots, \nonumber
\end{eqnarray}
where the infinitesimal function $a(\tau)$ is constructed out of
the parameters $a_n$ \begin{equation}\label{a}
a(\tau)=a_0+a_1\tau+a_2\tau^2+ +a_3\tau^3\ldots =
\sum_{n=0}^{\infty}a_n\tau^n.
\end{equation}
One can see from \p{Transtau} that the parameter $\tau$ transforms
precisely as the coordinate of the one-dimensional space under the
reparametrization. The parameters $x_0$ and $x_1$ transform
correspondingly as the dilaton and one-dimensional Cristoffel
symbol. In general the transformation rule for $x_n$ contains
$(n+1)$-st derivative of the infinitesimal parameter $a(\tau)$.

To make the connection with the physical models, it is natural to
consider all parameters $x_n, n=0,1,2,...$ in \p{coset1} as the
fields in one-dimensional space parametrized by the coordinate
$\tau$.

The conformal group $SL(2,R) \sim SU(1,1)$ in one dimension is a
three-parameters subgroup of \p{coset1}, namely the one generated
by $L_{-1}, L_0$ and $L_1$. Its group element is a product of
first two and last one multipliers in the expression \p{coset1}
\begin{equation}\label{coset2}
G_C=e^{i\tau L_{-1}} \cdot e^{ix_1L_1} \cdot e^{i{x_0}L_0}.
 \end{equation}
In other words, the $SL(2,R)$ group is embedded in the Virasoro
group in the simplest way by the conditions
\begin{equation}\label{Conditions}
 x_n=0,\quad n\geq 2
\end{equation}
The infinitesimal transformation function $a(\tau)$ \p{epsilon}
which conserves the conditions \p{Conditions} contains only three
parameters
\begin{equation}\label{3parameter}
  a(\tau)= a_0+a_1\tau +a_2\tau^2.
\end{equation}

It is convenient to introduce new variables playing the roles of
the coordinate and momentum of the particle
\begin{equation}\label{newVar}
  x=e^{x_0/2},\quad p=x_1 x,
\end{equation}
for which the conformal group infinitesimal transformations are
\begin{eqnarray}\label{Transtau3parameter}
\delta \tau &=&a(\tau),\\
\delta x&=&\frac{1}{2}\dot{a}(\tau)x,\label{TransX}\\
\delta {p}&=& -\frac{1}{2}\dot{a}(\tau)p+
\frac{1}{2}\ddot{a}(\tau)x,\label{TransP}
\end{eqnarray}
with $a(\tau)$ given in this case by the expression
\p{3parameter}. The corresponding finite transformations are
\begin{equation}\label{CT}
\tau'=\frac{a\tau+b}{c\tau+d},\quad x'=\frac{x}{c\tau+d},\quad
p'=(c\tau+d)p-cx,
\end{equation}
with parameters of the transformation constrained by the
unimodularity condition $ad-bc=1$.

\setcounter{equation}0
\section{The Application of Nonlinear Realizations of the
$SL(2,R)$ Group for the Actions Construction}
\subsection{The action integral for Conformal Mechanics}
The transformations \p{CT} form a symmetry group of the Conformal
Quantum Mechanics of \cite{AFF} with the action
\be\label{ActionAFF} S = {1\over 2} \int d\tau \left( \dot{x}^2 -
{\gamma\over x^2} \right). \ee As was shown in \cite{IKL1} (see
also \cite{ACP}) this action can be naturally described on the
language of invariant differential Cartan's form
\begin{equation}\label{CartanC}
\Omega_C=G_C^{-1} dG_C=\Omega_{-1}L_{-1}+\Omega_0 L_0+\Omega_1 L_1
\end{equation}
connected with the parametrization \p{coset2} of the conformal
group. The explicit calculations give
\begin{eqnarray}\label{Cbose} \Omega_{-1}&=&\frac{d\tau}{x^2}, \\
\Omega_0&=& \frac{dx-pd\tau}{x},\label{CartanC0}  \\
\Omega_1&=&xdp-pdx+p^2d\tau.\label{CartanC1}
\end{eqnarray}

All these differential forms are invariant under the
transformations \p{CT} and can be used for construction of an
invariant action. The simplest one is the linear in $\Omega$-forms
combination\cite{IKL1}
\begin{eqnarray}\label{BoseAction}
S&=&-\frac{1}{2}\int \Omega_1-\frac{\gamma}{2}\int\Omega_{-1}=\\
&&\frac{1}{2}\int d\tau \left(-x {\dot p}+p{\dot x} - p^2-
\frac{\gamma}{x^2} \right)~.\nonumber
\end{eqnarray}
The first term in this expression is appropriately normalized to
get the correct kinetic term. The parameter $\gamma$ plays the
role of cosmological constant in one dimension because
$\Omega_{-1}$, which corresponds to the translation generator
$L_{-1}$, is the differential $1$-form einbein.

The action \p{BoseAction} is a first order representation of the
action describing the Conformal Mechanics of De Alfaro, Fubini and
Furlan\cite{AFF}. Indeed, one can find  $p$ by solving its
equation of motion, insert it back in the lagrangian and get the
second order action \p{ActionAFF}.

\subsection{The action integral for Conformal Mechanics
with Additional Harmonic Potential} From the point of view of
underlying physics the action \p{ActionAFF} is not satisfactory
one, because the corresponding quantum mechanical Hamiltonian does
not have the ground state. The modification of this action with
the appealing spectrum of the energy was considered in \cite{AFF}.
It includes the additional harmonic oscillator term
\be\label{ActionAFFHarmonic} S_1 = {1\over 2} \int d\tau \left(
\dot{x}^2 -  {\gamma\over x^2}-\omega^2 x^2 \right). \ee Though
the action \p{ActionAFFHarmonic} contains the dimensional
parameter $\omega$, it is invariant under the transformations of
conformal group, realized by the more complicated expressions, as
we will see.

As we have already mentioned in the Introduction, the action
\p{ActionAFFHarmonic} can not be described in the framework of
nonlinear realizations of the $SL(2,R)$ group, parametrized as in
\p{coset2}. Instead, we will consider the embedding of this group
in the Virasoro group \p{coset1} by conditions different from the
simplest ones \p{Conditions}. The structure of the component
$\Omega_1^V$ in the Cartan's Omega-form connected with the
Virasoro group
\begin{equation}\label{CartanV}
\Omega_V=G^{-1} dG=\Omega_{-1}L_{-1}+\Omega_0 L_0+\Omega_1^V L_1+
\Omega_2^V L_2+\ldots
\end{equation}
may serve as a hint in the choice of the appropriate conditions.
The components $\Omega_{-1}$ and $\Omega_{0}$ coincide with the
corresponding components \p{Cbose} and \p{CartanC0}. Though the
components $\Omega_2^V, \Omega_3^V, \ldots$ depend in general on
all parameters $x_n$, the component $\Omega_1^V$
\begin{equation}\label{CartanV1}
\Omega_1^V=xdp-pdx+p^2d\tau-3x_2x^2d\tau
\end{equation}
depends in addition to the phase space variables $(x,p)$ only on
the parameter $x_2=x_2(\tau)$. So, the last term in the expression
\p{CartanV1} is the only difference of it with respect to the
corresponding expression \p{CartanC1} calculated for the
representation \p{coset1} of the $SL(2,R)$ group. Moreover, if we
take
\begin{equation}\label{x2omega}
x_2(\tau)=-\frac{1}{3}\omega^2,\quad \omega=const,
\end{equation}
we obtain exactly an oscillator-like term in the action
\begin{eqnarray}\label{OmegaAction}
  S&=&-\frac{1}{2}\int \Omega_1^V-\frac{\gamma}{2}\int\Omega_{-1}=\\
&&\frac{1}{2}\int d\tau \left(-x {\dot p}+p{\dot x} -
p^2-\omega^2x^2- \frac{\gamma}{x^2} \right)~,\nonumber
\end{eqnarray}
which coincides with the action $S_1$ \p{ActionAFFHarmonic} in the
second order form.

The component $\Omega_1^V$ \p{CartanV1} by construction is
invariant under the arbitrary infinitesimal transformations
\p{left1} of the Virasoro group if parameters $x, p,$ and $x_2$
transform according to the equations \p{TransX}, \p{TransP} and
\p{Trans2}. The consistency condition of this last transformation
law with the demand that $\omega=const$ can be written in the form
\begin{equation}\label{EqOmegaConst}
\stackrel{\ldots}{a}(\tau)+4\omega^2\dot{a}(\tau)=0.
\end{equation}
The solution of this differential equation gives
\begin{equation}\label{ParOmegaConst}
a(\tau)=a_0+a_1\sin(2\omega\tau)+a_2\cos(2\omega\tau).
\end{equation}
So, the action of Conformal Mechanics \p{ActionAFFHarmonic} with
additional oscillator-like potential is invariant under the three
parameter transformation \p{ParOmegaConst}.

\subsection{The Time Dependent Oscillator}
In general the variable $x_2(\tau)$ can be arbitrary function of
the time. Nevertheless, the treating it as a dynamical variable
leads to the trivial dynamics because, as one can easily see from
the expression \p{CartanV1}, it plays the role of a Lagrange
multiplier leading to the equation of motion $x^2=0$\footnote{If
the variable $x$ carries in addition some index $I$ -
$x\rightarrow x_I$ the situation drastically changes when this
index describes the vector representation of the rotation group of
the space-time with the signature $(D,2)$. In this case the action
is given by the sum of $D+2$ expressions \p{CartanV1} (with the
corresponding signs) and it describes the massless particle in $D$
- dimensional space-time\cite{P2} (the spinning particle if
instead of the Virasoro group one considers the reparametrization
group in the superspace $(1,N)$ with one bosonic and $N$ Grassmann
coordinates\cite{MAAP})} So, instead of being the constant as in a
previous Subsection, the parameter $x_2$ in a physical model can
be at most some {\em fixed} function $x_2(\tau)$. If we are
looking for the invariance transformations of the action
\p{OmegaAction} with the time dependent frequency $\omega^2(\tau)$
($x_2(\tau)=-\omega^2(\tau)/3$), it means that after the time
transformation \p{Transtau} $\tau\rightarrow \tau'=\tau+a(\tau)$
the functional dependence should remain the same:
$x_2(\tau)\rightarrow x_2(\tau'), \delta x_2(\tau)=a(\tau)\dot
x_2(\tau)$. The transformation law \p{Trans2} leads then to the
equation for the infinitesimal parameter $a(\tau)$
\begin{equation}\label{EqOmegaTime}
\stackrel{\ldots}{a}(\tau)+4\omega^2(\tau)\dot{a}(\tau)+
2\frac{d}{d\tau}{(\omega^2(\tau))}{a}(\tau)=0.
\end{equation}
This differential equation of the third order with the time
dependent coefficients has the solution in the form\cite{K}
\begin{equation}\label{Solution}
  a(\tau)=C_1 u_1^2+C_2 u_1u_2+C_3 u_2^2,
\end{equation}
where $C_1, C_2, C_3$ are three infinitesimal constants and
functions $u_1(\tau), u_2(\tau)$ form the fundamental system of
solutions of auxiliary equation
\begin{equation}\label{Aux}
\ddot{u}(\tau)+\omega^2(\tau){u}(\tau)=0.
\end{equation}
For the time independent $\omega$ this solution reproduces the
ones given by \p{ParOmegaConst}. For different particular forms of
the $\omega^2(\tau)$ the equation \p{EqOmegaTime} becomes, for
example, the Lame, Matieu, Hill etc. equations \cite{K}, each of
which play the very important role in the physics.

So, the solution \p{Solution} of the equation \p{EqOmegaTime}
describe the invariance transformations of the action for the Time
Dependent Oscillator with the frequency $\omega(\tau)$.

The very important for the Time Dependent Oscillator model is the
question about its possible connection with other solvable systems
like Harmonic Oscillator or even with the free particle. Such
connection can in principle lead to the construction of exact
solutions of the Schr\"{o}dinger equation for the Time Dependent
Oscillator starting from the solutions of these simpler systems.
The example of such connection was given in \cite{AFF} where the
transformation from the system with vanishing frequency
\p{ActionAFF} to the one with constant $\omega$
\p{ActionAFFHarmonic} was given. This transformation inevitably
includes the transformation of the time.

To construct the generalizations of this transformation let us
consider the most general finite transformation of the Virasoro
group\cite{ACP}
\begin{eqnarray}\label{TransGeneral1}
\tau &\rightarrow&\tau'= f(\tau),\\
x(\tau)&\rightarrow&x'(\tau')=
({\dot f}(\tau))^{1/2}x(\tau),\label{TransGeneral2}\\
p(\tau)&\rightarrow&p'(\tau')=\frac{1}{({\dot
f}(\tau))^{1/2}}p(\tau)+ \frac{{\ddot f}(\tau)}{2({\dot
f}(\tau))^{3/2}}x(\tau),
\label{TransGeneral3}\\
x_2(\tau)&\rightarrow&x'_2(\tau')= \frac{1}{({\dot
f}(\tau))^{2}}x_2(\tau)+ \frac{1}{2}\frac{d}{d\tau}
\left(\frac{\ddot f(\tau)}{\dot f(\tau)}\right)-
\frac{1}{4}\left(\frac{\ddot f(\tau)}{\dot f(\tau)}\right)^2.
\label{TransGeneral4}
\end{eqnarray}
If we start with $x_2(\tau)=0$, i.e. in the absence of the
oscillator like potential, the frequency of the induced harmonic
term in the new time will be given by the expression
\begin{equation}\label{Freq}
\omega^2(\tau)=\frac{1}{2}\frac{d}{d\tau} \left(\frac{\ddot
f(\tau)}{\dot f(\tau)}\right)- \frac{1}{4}\left(\frac{\ddot
f(\tau)}{\dot f(\tau)}\right)^2,
\end{equation}
which can be recognized as the Schwarz derivative of the
transformed time over the old ones (see also \cite{CCKM}-\cite{ACKZ}
where
$x_2(\tau)$ was considered as an external field).
One can rewrite the equation
\p{Freq} in the more familiar form
\begin{equation}\label{TintoTau}
 \left(\frac{d^2}{d\tau^2}+
 \omega^2(\tau)\right)\frac{1}{\sqrt{\dot f(\tau)}}=0.
\end{equation}
So, the solution $f(\tau)$ of the equation \p{TintoTau} gives the
transformation rules \p{TransGeneral1}-\p{TransGeneral4} between
the two systems - the one without oscillator-like potential and
the others having in addition to the conformal potential $\sim
1/x^2$ the term $\omega^2(\tau)x^2$. One should note also that the
conformal potential term in the action is by itself invariant
under the arbitrary finite transformations
\p{TransGeneral1}-\p{TransGeneral2}.

\section{Conclusions}
In this paper we applied the methods of nonlinear realizations
approach for construction of the actions of Conformal Quantum
Mechanics, as well, as the action of the Time Dependent
Oscillator. We have shown that both the actions are invariant
under the three parameter transformations which are nontrivially
embedded in the Virasoro group. We described also different types
of transformations belonging to the Virasoro group and making the
connections of different systems each with other. In particular,
we obtain such transformation between the free motion Hamiltonian
and Hamiltonian of the Time dependent Oscillator. It would be
interesting to carry up the analogous considerations in the more
complicated theories, such as $N=2$ and $N=4$ SuperConformal
Quantum Mechanics.

\section*{Acknowledgements}

A.P. thanks the members of Graduate School and University Center,
the City University of New York, where the essential part of this
work was done. The work of A.P.  was partially supported by INTAS
grant, project No 00-00254 and RFBR grant, project No 03-02-17440.

\end{document}